\documentclass[12pt]{article}
\usepackage{amsmath, amssymb, slashed, epsf, color}
\usepackage[dvipdfmx]{graphicx}

\begin{document}
\begin{titlepage}
\begin{center}

\hfill IPMU-14-0160 \\
\hfill ICRR-Reprot-687-2014-13

\vspace{2.0cm}
{\large\bf
C$\nu$B absorption line in the neutrino spectrum at IceCube
}

\vspace{2.0cm}
{\bf Masahiro Ibe}$^{(a,b)}$
and {\bf Kunio Kaneta}$^{(a)}$\\
\vspace{1.5cm}
{\it
$^{(a)}${ICRR, University of Tokyo, Kashiwa, Chiba 277-8582, Japan}\\
$^{(b)}${Kavli IPMU (WPI), University of Tokyo, Kashiwa, Chiba 277-8583, Japan}
}

\vspace{2.5cm}
\abstract{
The IceCube experiment has recently reported high energy neutrino spectrum between TeV$-$PeV scale.
The observed neutrino flux can be as a whole well fitted by a simple power-law of the neutrino energy $E_\nu$, $E_\nu^{-\gamma_\nu}$ ($\gamma_\nu \simeq 2$).
As a notable feature of the spectrum, however, it has a gap between 500\,TeV and 1\,PeV.
Although the existence of the gap in the neutrino spectrum is not statistically significant at this point,
it is very enticing to ask whether it might hint some physics beyond the Standard Model.
In this paper, we investigate a possibility that the gap can be interpreted as an absorption 
line in the power-law spectrum 
by the cosmic neutrino background through a new resonance in the MeV range. 
We also show that the absorption line has rich information about not only the MeV scale new particle 
but also the neutrino masses as well as the distances to the astrophysical sources of the high energy neutrinos.
Viable models to achieve this possibility are also discussed.
}

\end{center}
\end{titlepage}
\setcounter{footnote}{0}

\section{Introduction}

The IceCube experiment has recently reported high energy neutrinos considered to be coming from extraterrestrial source since those observed events are significantly large compared to the atmospheric neutrino background~\cite{Achterberg:2006md,Aartsen:2014gkd}.
Such high energetic neutrinos are expected to come from, for example, the photo-pion production such as 
$\gamma p\to \Delta \to \pi^+ X$ followed by the pion decay, $\pi^+ \to \nu_\mu(\mu^+\to \nu_e \bar\nu_\mu e^+)$ which produces the neutrinos of the flavor composition with $\nu_e:\nu_\mu:\nu_\tau=1:2:0$ while it becomes $1:1:1$ after traveling from some extraterrestrial source\footnote{The flavor oscillation of the neutrino being the energy $E_\nu$ would take place after traveling the distance $L\sim 2E_\nu/\Delta m_{ij}^2$ where the mass difference is defined by $\Delta m_{ij}^2\equiv m_{\nu_i}^2-m_{\nu_j}^2$ for the mass eigenstate of neutrinos $\nu_i$, and if we take $\Delta m_{21}^2\sim 10^{-3}~{\rm eV^2}$ and $E_\nu\sim 10^{6}~{\rm GeV}$, the distance becomes $L\sim10^{-10}~{\rm Mpc}$ which is small enough even if some astrophysical neutrino source locates within the intergalactic scale~\cite{Lunardini:2000fy,Athar:2000yw}.}. 
The cosmogenic neutrino flux, however, peaks at around $O(1)$\,EeV for $\gamma$ being the
cosmic microwave background (CMB) and is difficult to explain the observed neutrino flux in the sub-PeV 
region~\cite{Yoshida:1993pt,Takami:2007pp}. 
As other possibilities, there are many candidates to explain the events around the sub-PeV to the PeV region 
by the high energetic cosmic-ray sources inside our galaxy such as the supernova Remnants 
(SNR)~\cite{Villante:2008qg,Murase:2013ffa} and the pulsar wind nebulae (PWN)~\cite{Bednarek:2003cv}
as well as the extra-galactic sources such as the gamma ray bursts (GRB)~\cite{Asano:2014nba,Dado:2014mea}, 
the active galactic nuclei (AGN)~\cite{Stecker:2013fxa}, and the star forming galaxies~\cite{Tamborra:2014xia}
(see also Refs.~\cite{Learned:2000sw,Dermer:2006xt,Cholis:2012kq,Murase:2013rfa,Anchordoqui:2013dnh} and references therein).
More ambitious explanations by physics beyond the SM such as 
decaying dark matter or new interactions of neutrino have also been 
discussed~\cite{Feldstein:2013kka,Ema:2013nda,Esmaili:2013gha,Ng:2014pca,Ioka:2014kca,Zavala:2014dla}.

As a current status of the observed neutrino flux, on the other hand, it is as a whole well fitted by
a simple power-law $E_\nu^{-\gamma_\nu}$ ($\gamma_\nu \simeq 2$), in the sub-PeV to the PeV range,
where $E_\nu$ is the observed neutrino energy.
This power spectrum is vaguely supported by the source spectrum of the cosmic ray proton
accelerated by the first order Fermi acceleration mechanism.
As a notable feature of the spectrum, however, it has a gap between 500\,TeV and 1\,PeV.
Although the existence of the gap in the observed neutrino spectrum is not statistically significant at this point (see e.g. \cite{Chen:2013dza}),
it is very enticing to ask whether it might hint some physics beyond the standard model (SM).

In this paper, we investigate a possibility that the gap in the power-law spectrum 
can be interpreted as an absorption line by the cosmic neutrino background (C$\nu$B) 
through a new resonance with a mass in the MeV range. 
We also show that the neutrino absorption line has rich information about not only the MeV scale new particle 
but also the neutrino masses as well as the distances to the astrophysical sources of the neutrinos.
Viable models to achieve this possibility are also discussed.

\section{New particle and resonant absorption}
Let us discuss whether it is possible to interpret the null event regions at the sub-PeV neutrinos 
as the C$\nu$B absorption line in the single power law spectrum of $E_\nu^{-\gamma_\nu}$ with $\gamma_\nu = 2$.
In the SM,  there is no appropriate interactions which shows an absorption line at the sub-PeV region.
As we will see shortly, however, such an absorption line interpretation becomes possible 
by introducing a new resonance appearing in the $s$-channel neutrino-(anti)neutrino scattering. 

The C$\nu$B is a remnant of the primordial plasma reheated after the inflation, 
and the temperature of the C$\nu$B is predicted to be $T_\nu\simeq 1.96~{\rm K}\simeq 1.69\times 10^{-4}$\,eV.
From this temperature, the neutrino number density is given by $n_\nu\simeq 56~{\rm cm^{-3}}$ for each flavor.
When the high energy neutrinos  accelerated by some astrophysical source  collide
with the C$\nu$B of the masses larger than 
$T_\nu$, the situation is almost the same as the collision with a fixed target in the laboratory frame. 
In this case, the center-of mass energy is given by $\sqrt{2m_\nu E_\nu}$, where $m_\nu$ denotes 
the mass of the target neutrino in the  C$\nu$B.
Thus, if the mass of a new particle, $M_s$, appearing in the $s$-channel neutrino collisions 
is around $M_s \simeq \sqrt{2m_\nu E_\nu}$, the injected neutrinos of $E_\nu$ are resonantly
scattered by the C$\nu$B, which leads to the ``absorption line" in the neutrino spectrum. 
For example, $E_\nu\sim 1~{\rm PeV}$ neutrino absorption predicts a new particle in the mass around 
$M_s\sim 10~{\rm MeV}$ if we take the neutrino mass  $m_\nu=0.1~{\rm eV}$.

Before introducing a new particle, however, let us first examine what 
is expected on the neutrino spectrum in the SM.
There, most of the cosmic-ray neutrinos accelerated by some astrophysical sources
are expected to penetrate astrophysical/cosmological distances  since they interact 
with materials very weakly.
As the neutrinos are traveling in the distance, 
the most relevant target material is the C$\nu$B since 
it is as abundant as the CMB while it has larger interaction rates with the neutrino flux than the CMB.
In the SM, the neutrinos interact with themselves via the electroweak interactions, 
where the relevant processes are $\nu_l \bar\nu_{l,{\rm C}\nu{\rm B}}\to f\bar f~(f=\nu_l, l, q, \cdots)$, 
$\nu_l \bar\nu_{l',{\rm C}\nu{\rm B}}\to \nu_l \bar\nu_{l'}, l\bar l'~(l \neq l')$, 
and $\nu_l \nu_{l',{\rm C}\nu{\rm B}}\to \nu_l \nu_{l'}$.
The cross sections of the SM processes are given in, for example, Refs.~\cite{Ema:2013nda,Roulet:1992pz}.
Since some of them can be enhanced via $s$-channel $Z$-boson exchanges 
at the energy of $Z$ boson mass, neutrino absorption may occur for the energy of neutrino flux around 
$E_\nu=M_Z^2/(2m_\nu)\sim10^{13}~{\rm GeV}$. 
This absorption line is far above the energy range of the recently observed neutrinos,
and hence, we cannot attribute the null event regions in the IceCube spectrum
to the absorption line in the SM.
The occurrence of such an absorption feature by $Z$-boson is known as "Weiler mechanism", which has been studied in Refs.~\cite{Weiler:1982qy,Weiler:1983xx,Weiler:1997sh,Yoshida:1994ci,Yoshida:1996ie}.
Related topics have been also studied in \cite{Fodor:2001qy,Eberle:2004ua,Barenboim:2004di}.

Now, let us introduce a new light particle to make an absorption line at around the sub-PeV range
in the neutrino spectrum.
The situation is similar to the  $Z$-boson resonance, while the new particle coupling to the neutrinos 
are predicted to be around MeV scale in our case as mentioned above.
Suppose that the new scalar particle $s$ with a mass $M_s$ couples to the neutrinos by
\begin{equation}
	{\cal L}_{s-\nu}=g s\bar\nu_i \nu_j
	\label{NSI}
\end{equation}
with coupling $g$ where we assume that the coupling is flavor universal for simplicity.
Here, we do not specify whether the neutrino is the Dirac type or the Majorana type.
One caution is, however, that if the above interaction is the Yukawa interaction
between the left-handed and the right-handed neutrinos of the Dirac neutrino,
the right-handed neutrinos are copiously produced in the early universe through 
this interaction.
Such a possibility is severely restricted by the constraints on the effective number 
of neutrinos, $N_{\rm eff} = 3.02\pm 0.27$ from the big-bang nucleosynthesis and the CMB observations~\cite{Kirilova:2014ipa} which eventually leads to a constraint on the coupling constant;
\begin{eqnarray}
g \lesssim ( M_s/ M_{\rm PL})^{1/4}\ .
\end{eqnarray}
Here $M_{\rm PL}$ denotes the reduced Planck mass $M_{\rm PL} \simeq 2.4 \times 10^{18}$\,GeV.
Since we will use rather sizeable coupling constants, we find that the only possible interactions are
\begin{equation}
	{\cal L}_{s-\nu} = 
\left\{
\begin{array}{cc}
g s\nu_{Li} \nu_{Lj}\ , & \mbox{(Majorana, Dirac)} \ ,
 \\
g s\bar N_{Ri} \bar N_{Rj}\ , & \mbox{(Dirac)}\ ,
\end{array}
\right.
	\label{NSI2}
\end{equation}
where $\nu_L$ and $\bar {N}_R$ denotes the left-handed neutrinos and the right-handed neutrinos.
Flavor dependence of the coupling as well as the consistency with the electroweak theory 
will be discussed in next section.
\begin{figure}
\begin{center}
	\includegraphics[scale=.7]{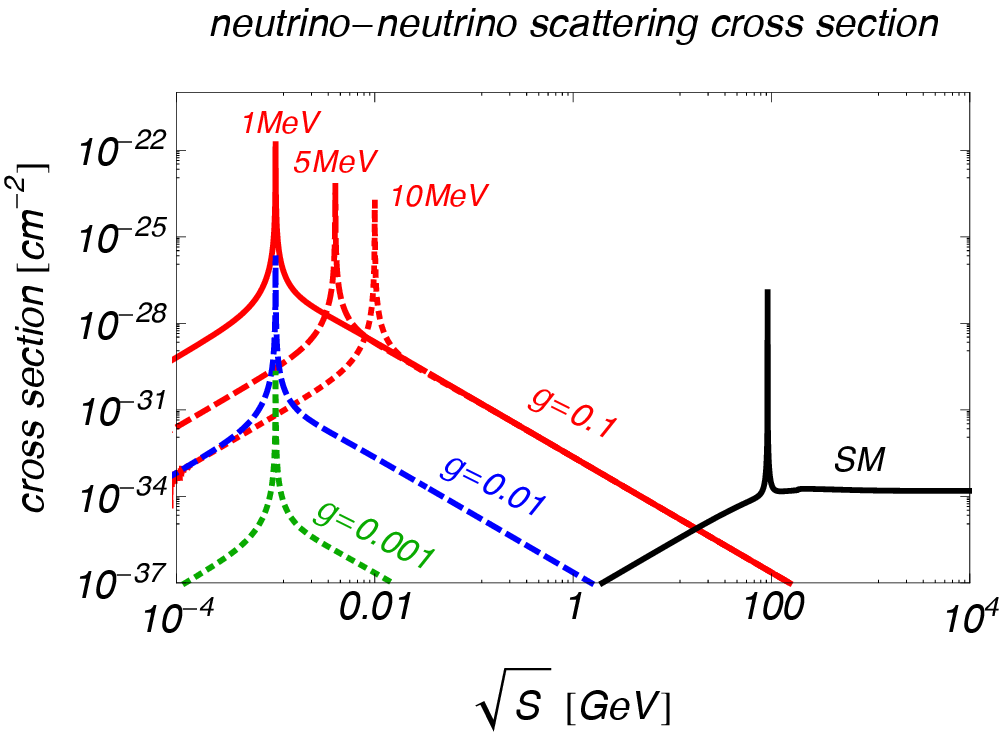}
	\includegraphics[scale=.55]{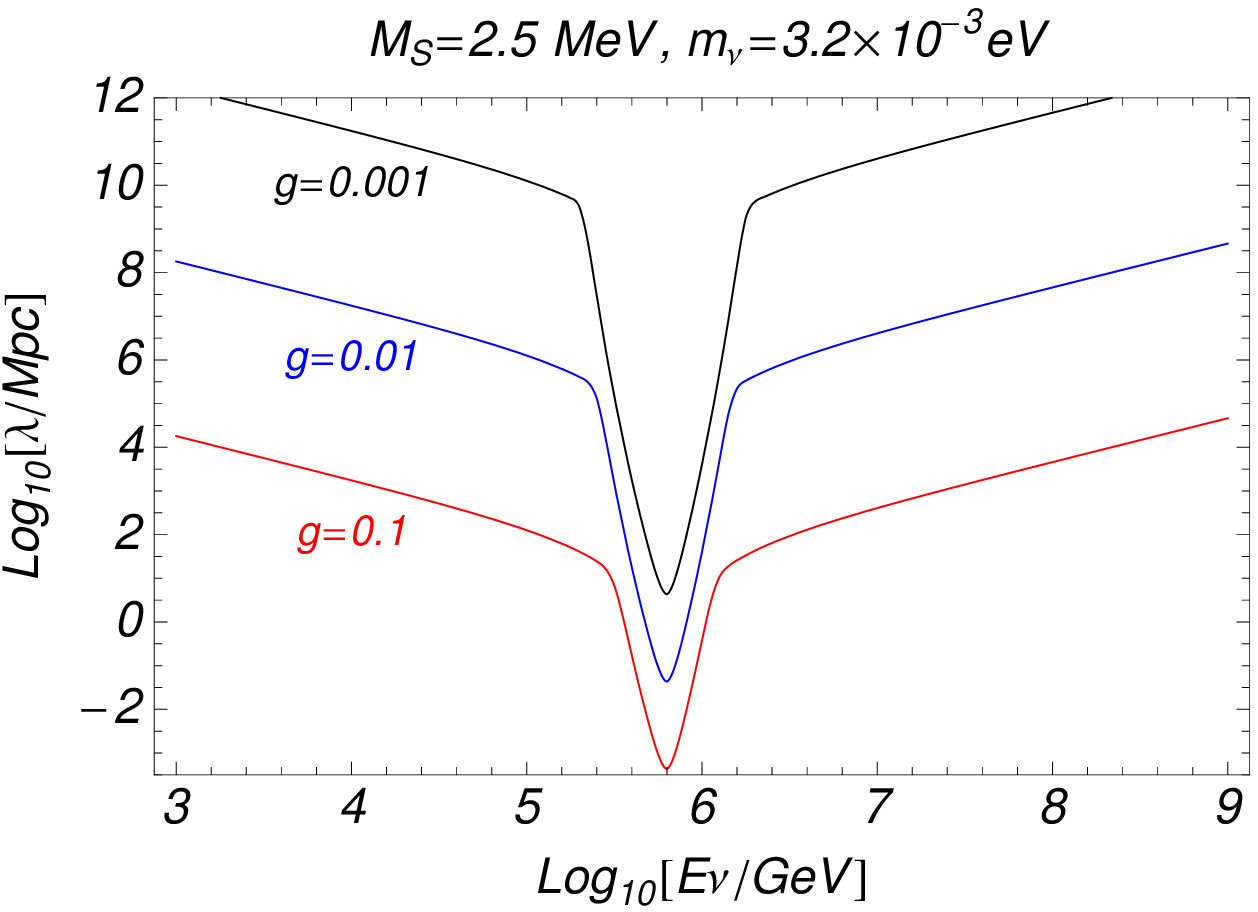}
	\caption{\sl \small {\bf Left panel:} The neutrino-(anti)neutrino scattering cross sections for the center of mass energy $\sqrt{S}$. The black solid line is the SM. The red solid, the red dashed, the red dotted, the blue dashed, and the green dotted lines depict the contributions from the interaction of Eq.(\ref{NSI}) for several parameter samples 
	of the coupling $g$ and the scalar particle mass.
	{\bf Right panel:} The neutrino mean free path $\lambda$ as function of the energy of the neutrino flux.}
	\label{fig1}
\end{center}
\end{figure}

The neutrino-(anti)neutrino scattering cross section $\sigma_{\nu\nu}(S)$ are evaluated as shown in the left panel of Fig.~\ref{fig1}.
The black solid line shows the SM cross section with the resonance at the $Z$ boson pole.
The parameters $(g, M_s)$ for the cross section by the new resonance are as indicated.
The highest value of the cross section at $S=M_s^2$ is determined by the decay width of $s$ given by
\begin{equation}
	\Gamma_s=N_\nu\frac{g^2}{16\pi}M_s
	\left[
		1-\frac{2m_\nu^2}{M_s^2}
	\right]
	\left[
		1-\frac{4m_\nu^2}{M_s^2}
	\right]^{1/2},
\end{equation}
where $s$ is assumed to decay into $N_\nu$ neutrinos, and consequently the peak of the cross section is $\sigma_{\nu\nu}(S=M_s^2)\simeq 16\pi/(N_\nu^2 M_s^2)$.

The neutrino mean free path (MFP) $\lambda$ is an important quantity to evaluate how far the neutrino traveling distance is.
The MFP is defined by
\begin{eqnarray}
	\lambda(E_\nu) &=& 
	\left[
		\int \frac{d^3p}{(2\pi)^3} \sigma_{\nu\nu}(E_\nu,p)f_{\nu}(p)
	\right]^{-1},
\end{eqnarray}
where $f_{\nu_i}(p)$ is the C$\nu$B distribution function given by $f_{\nu_i}(p)=[\exp (|\vec p|/T_\nu)+1]^{-1}$.
Examples of the MFP are shown in the right panel of Fig.~\ref{fig1} where $M_s$ and $m_\nu$ is set to $M_s=2.5~{\rm MeV}$ and $m_\nu=3.2\times10^{-3}~{\rm eV}$, respectively.
The black, the  blue, and the red solid lines respectively show the case of $g=0.001,~0.01$, and $0.1$.
If the traveling distance of the neutrinos is below the lines, the neutrino flux at a corresponding energy can not reach to the Earth.
In most of energy region except for the resonance region, 
the relative magnitude among those lines is determined by the magnitude of the coupling, for example,
the MFP for the case of $g=0.1$ is four digits smaller than the case of $g=0.01$ 
since the cross section is proportional to $g^4$.
As indicated by the peak of the cross section, the bump structure of the MFP reflects the resonance of the singlet scalar.
Around the resonance region, the cross section is changing with a strength proportional to $g^2$, and thus the relative difference among MFPs is two digits magnitude.%
\footnote{At an energy near the resonance, the cross section behaves $\sigma(S=M_s^2+M_s\Gamma_s)\simeq \sigma(M_s^2)+g^2/M_s^2$.}

\begin{figure}
\begin{center}
	\includegraphics[scale=.5]{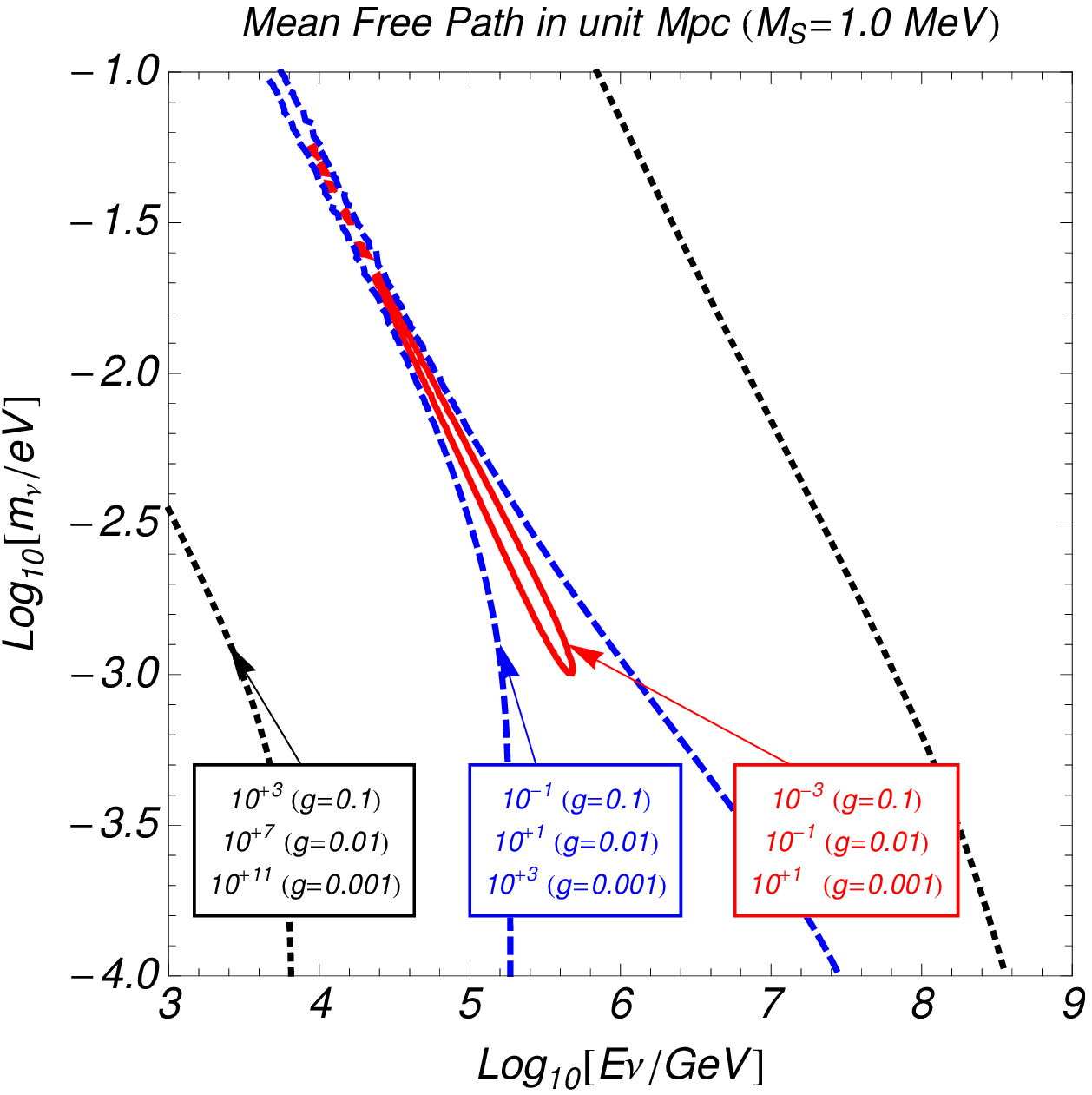}
	\includegraphics[scale=.5]{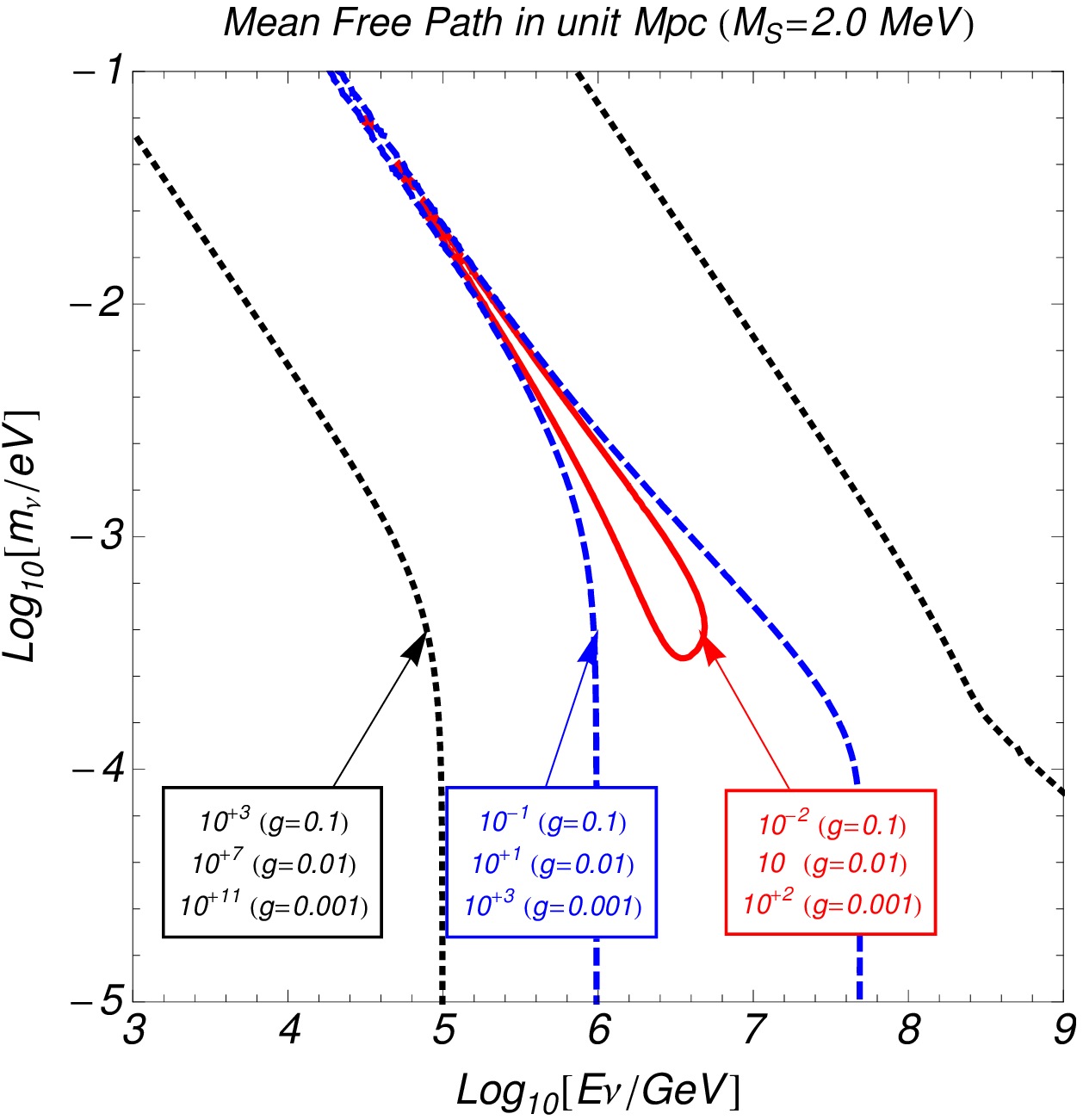}\\
	\includegraphics[scale=.5]{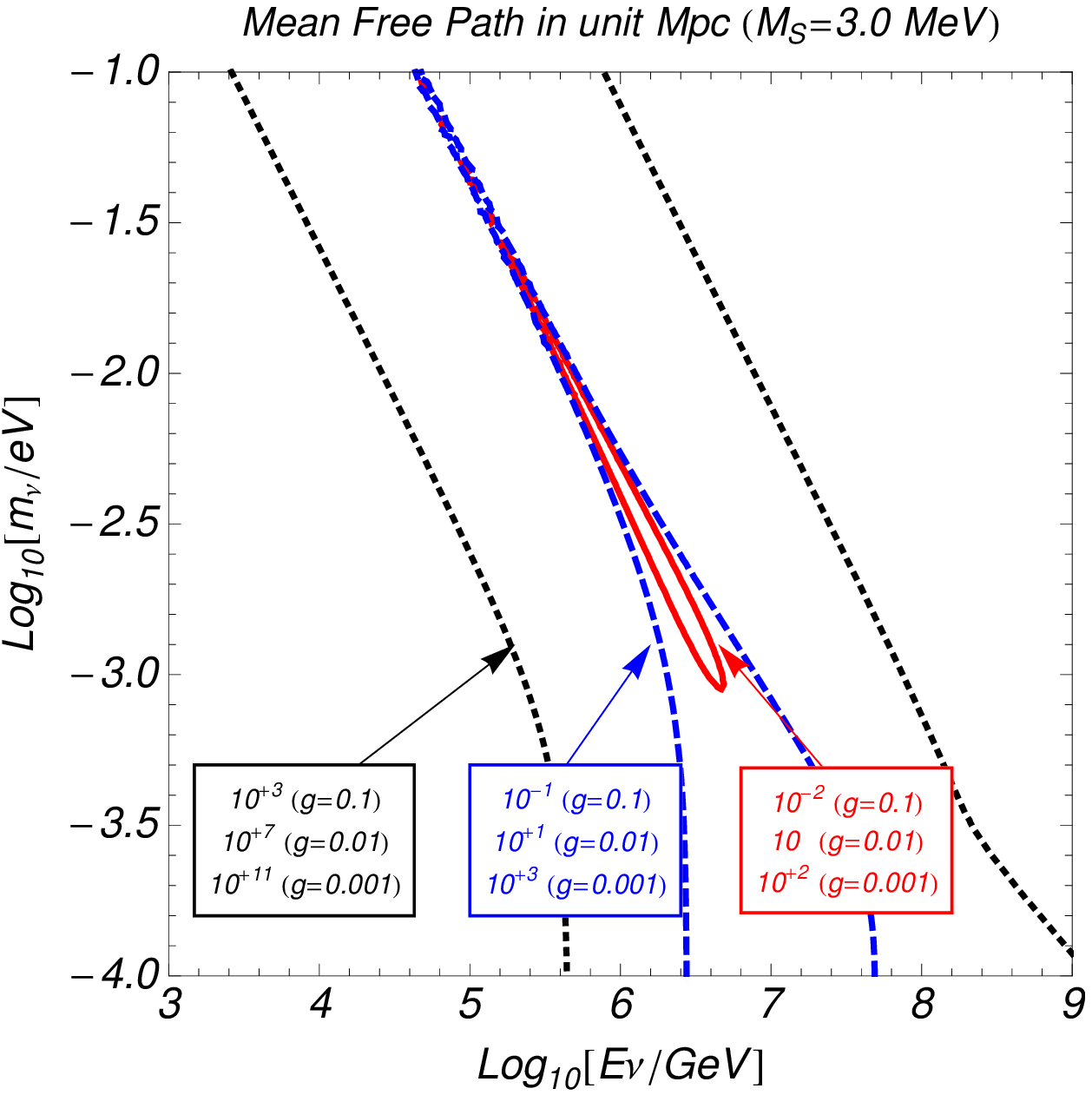}
	\includegraphics[scale=.5]{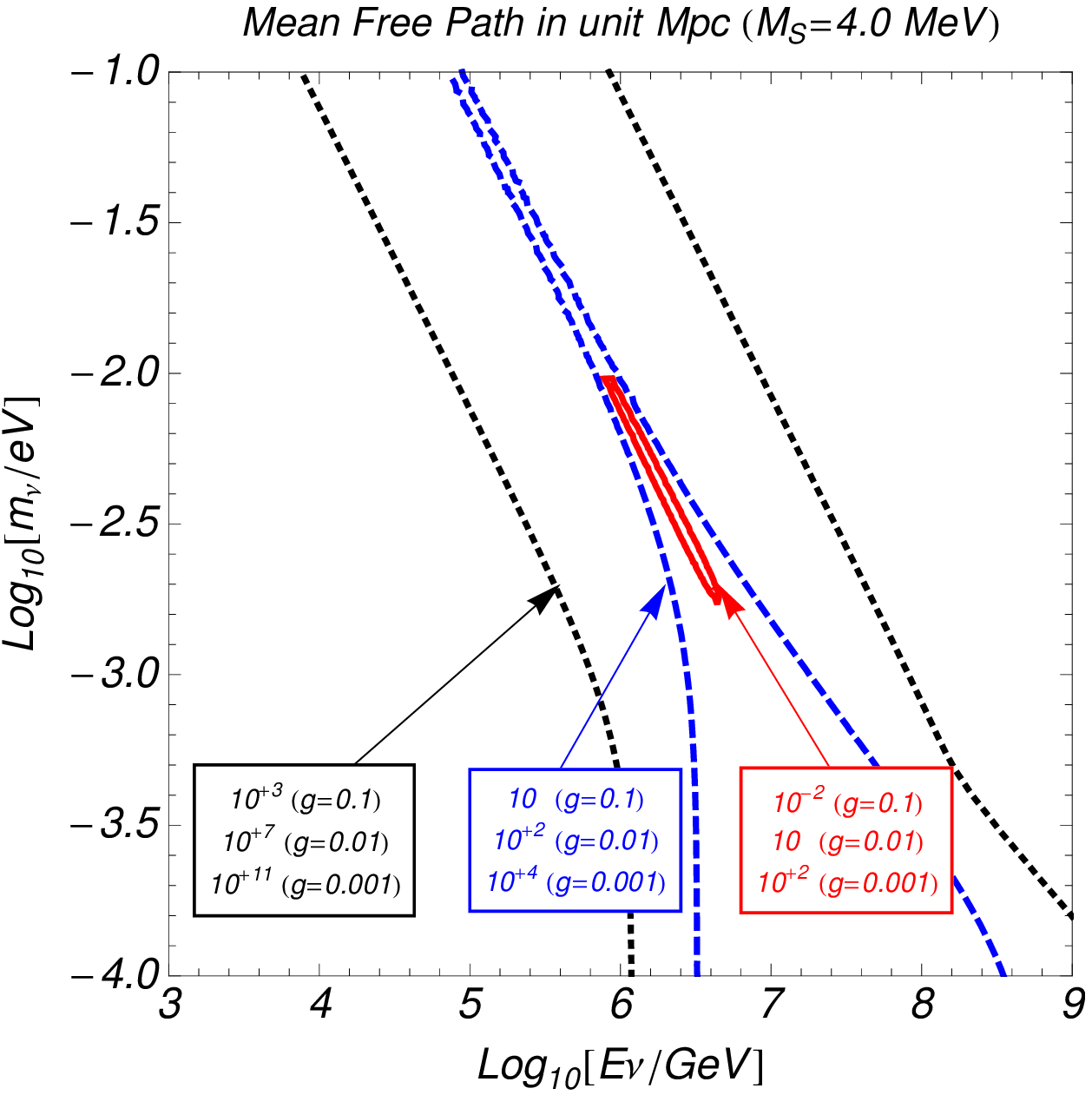}
	\caption{\sl\small The neutrino mean free path for various neutrino masses. 
	The numbers shown in the boxes are the mean free path in the unit {\rm Mpc} for each couplings. 
	The scalar boson mass is set to be $M_s=1~{\rm MeV},~2~{\rm MeV},~3~{\rm MeV}$ and $4~{\rm MeV}$ 
	in the upper left, upper right, bottom left and bottom right panels, respectively.}
	\label{fig2}
\end{center}
\end{figure}

Notably, the neutrino masses (of the C$\nu$B) are also an important parameter to determine the neutrino MFP.
Since the MFP is given by the overlap between the neutrino scattering cross section and the distribution function of 
the C$\nu$B,  it is sensitive to the neutrino mass through the center of the mass energy,
\begin{eqnarray}
 S \simeq 2 E_\nu \left( \sqrt{m_{\nu,{\rm C}\nu{\rm B}}^2+p_\nu^2} - p_\nu \cos\theta\right)\ ,
\end{eqnarray}
where $\cos\theta$ denotes the scattering angle and the typical value of $p_\nu$ is $O(T_\nu)$.
It should be noted that, $S$ becomes insensitive to $p_\nu$ and is solely determined by $E_\nu$
for $m_{\nu,{\rm C}\nu{\rm B}} \gg T_\nu$, while it takes wide range for $m_{\nu,{\rm C}\nu{\rm B}} \ll T_\nu$
due to the $p_\nu$ contribution.
Therefore the MFP becomes a sharp function of $E_\nu$ for $m_{\nu,{\rm C}\nu{\rm B}} \gg T_\nu$,
since $S\simeq M_s$ is achieved only for a particular value of $E_\nu$. 
On the contrary, the MFP becomes a broad function of $E_\nu$ for $m_{\nu,{\rm C}\nu{\rm B}} \gg T_\nu$
since wide range of $E_\nu$ can achieve $S\simeq M_s$.
The neutrino mass dependence of the MFP is shown in Fig.~\ref{fig2} which shows the contours of the MFP 
for a scalar boson mass of $M_s=1~{\rm MeV},~2~{\rm MeV},~3~{\rm MeV}$ and $4~{\rm MeV}$ with $g=0.1,~0.01$ and $0.001$ for each case.
The shortest MFP tends to be at a higher $E_\nu$ for a larger $M_s$.
In contrast, a heavier neutrino mass makes the shortest MFP be at lower $E_\nu$ since required $E_\nu$ 
to reach $\sqrt{S}=M_s$ becomes small for heavier $m_\nu$.
The figure indicates that the absorption line at around the sub-PeV region 
can be realized for $M_s \simeq 1$--$3$\,MeV, $m_{\nu}= 10^{-(2-2.5)}$ for 
the neutrino sources at the distance of $O(1)$\,Mpc. 

Before closing this section, let us comment on the traveling distance of the high energy neutrinos.
As mentioned in the introduction, there are several candidates for astrophysical source of the high energy neutrinos.
One of the promising candidates is the SNRs which locate typically 
$O(1$--$10)~{\rm kpc}$ far from the Earth.
The SNRs originated neutrinos are almost left-handed state even if they are massive since the neutrino 
has much higher energy than the neutrino mass.
Therefore, the absorption line scenario requires a new interaction involving left-handed neutrino 
so that the cosmic-ray neutrino scatter with the C$\nu$B, and the interaction should be strong enough 
to make the MFP shorter than $O(1$--$10$) kpc scale.
Other intriguing sources are the GRBs whose distances are $O(1)~{\rm Gpc}$ from the Earth.
In this case, an necessary coupling constant becomes smaller since the required MFP is longer than the case of SNRs.

\section{Viable models}
In the rest of this paper, we discuss viable models which is behind the effective theory considered in Eq.\,(\ref{NSI}).
So far, there have been many intriguing models in which neutrinos are interacting with new particles, 
for example, the Majoron models~\cite{Chikashige:1980ui,Gelmini:1980re}, 
the neutrinophilic Higgs models~\cite{Ma:2000cc},%
\footnote{We use the term "neutrinophilic" coined in Ref.~\cite{Haba:2011ra}.}
 the triplet Higgs models~\cite{Chanowitz:1985ug}.
However, straightforward adaptations of those models to our mechanism suffer from cosmological constraints 
and the constraints from the light meson rare decays since a rather large coupling of the neutrino interaction is required
for our purpose.%
\footnote{For the singlet Majoron model\,\cite{Chikashige:1980ui}, the resultant coupling between
the Majoron and the left-handed neutrinos are highly suppressed to achieve a light neutrinos.
The Majoron model appearing from the triplet Higgs may have a sizeable coupling 
to the left-handed neutrinos, although the model does not work for our purpose as we will comment later.}

\subsection{Inverse seesaw model with a neutrinophilic scalar doublet}
At first, let us examine a model where a neutrinophilic scalar doublet $h_N$ 
where, in addition to the usual right-handed neutrino $\bar N_R$, we also introduce
additional neutrinos $N_N$ which couple not to the Higgs doublet $h$  but only to $h_N$;
\begin{eqnarray}
	{\cal L} \supset g h_N l N_N + y h l \bar N_R + M \bar N_R N_N + m N_N N_N \ .
	\label{L2}
\end{eqnarray}
Here, $l$ denotes the lepton doublet in the SM,  $g$ and $y$ denote the dimensionless coupling constants, 
and $m$ and $M$ are the mass parameters.
In this model, we impose charges of the lepton number $L$ and the discrete symmetry $Z_2$ as shown in Table~\ref{tab1}.
Due to  these symmetries, the doublet scalar $h_N$ possesses the neutrinophilic  nature.
The last three terms induce the tiny neutrino mass, and by assuming $\langle{h_N}\rangle = 0 $,
the neutrino mass is given by $m_\nu\simeq y^2v^2(m/M^2)$ by the inverse seesaw mechanism~\cite{Deppisch:2004fa}. 
Here, $v$ is the vacuum expectation value (VEV) of the Higgs doublet,  
and $m\ll  y v\ll M$ is assumed.
The smallness of the neutrino mass is achieved by assuming that the 
lepton-number violating mass parameter $m$ is highly suppressed.
The other neutrinos than the three active neutrinos have masses of $O(M)$.

Let us emphasize the difference from the conventional model of the neutrinophilic Higgs doublet.
In the conventional neutrinophilic model, the neutrino masses are generated by the VEV of 
$h_N$, and hence, the neutrinos obtain the Dirac neutrino mass.
As discussed in the previous section, however, the Yukawa coupling between the left-handed and the right-handed neutrinos are severely restricted.
To avoid this problem, we separate the mass generation and the neutrino interaction by evoking the inverse seesaw mechanism.
As a result of the inverse seesaw mechanism, the Majorana neutrino masses and the effective coupling between $h_N$ and the left-handed neutrinos are simultaneously generated.

\begin{table}[ttbp]
\begin{center}
\begin{tabular}{c|ccccc||cc}
	    & $l$ & $N_N$ & $\bar N_R$ & $h$ & $h_N$ & $m$ & $M$\\ \hline\hline
	$L$ & $+1$ & $+1$ & $-1$ & $0$ & $-2$& $ -2 $ & $0$\\
	$Z_2$ & $+$ & $+$ & $-$ & $+$ & $-$ & $0$ & $-$ 
\end{tabular}
\caption{\sl \small Charge assignment of the model of Eq.~(\ref{L2}).
Here we also show the charge assignments of the mass parameters as spurious fields.}
\label{tab1}
\end{center}
\end{table}

Under the above symmetries, the scalar potential is given by
\begin{eqnarray}
	V &=& -\mu_h^2 |h|^2 + \lambda(h^\dagger h)^2 + \mu_N^2 |h_N|^2 + \lambda_1 (h_N^\dagger h_N)^2 \nonumber\\
	&& + \lambda_2 |h|^2|h_N|^2 - \lambda_3 |h^\dagger h_N + h.c.|^2
\end{eqnarray}
where $h_N$ does not acquire a VEV,%
\footnote{
Suppose that $m$ in Eq.~(\ref{L2}) are spurions of explicit breaking of the lepton number, it has the charge $L=+2$.
Therefore, $h_N$ and $h$ mix with each other in a form of $m^* h_N\leftrightarrow h$, whose mixing is of order $m/M\ll1$ via one-loop diagram, and consequently the VEV of $h_N$ is negligible 
and the contribution to the neutrino masses are also suppressed.
} 
and parameters $\mu_h^2,~\mu_N^2,~\lambda,~\lambda_1,~\lambda_2$ and $\lambda_3$ are defined as positive values.%
\footnote{
It should be noted that the last term of the potential explicitly breaks $L$ symmetry into $Z_4$ symmetry.
Consequently, Leptogenesis does not work since $B-L$ asymmetry is washed out by the explicit breaking term.
Therefore, alternative Baryogenesis scenario is necessary to generate the baryon asymmetry without sphaleron process.
}
We can estimate the scalar mass spectrum of $h_N$ by decomposing into $h_N=(h_N^0+iA^0,h_N^-)^T$ in which $h_N^0,~A^0$ and $h_N^-$ are neutral CP-even, neutral CP-odd, charged scalars, respectively.
They
acquire masses from the third, the fifth and  the 
last terms, meanwhile, only $h_N^0$ has an additional mass from the sixth term;%
\footnote{Generically, the term proportional to $\lambda_3$ contains two independent terms
which are allowed any symmetries than the custodial symmetry.
In our model, to evaded the constraints from the electroweak precisions, 
we fine-tune the potential so that  the scalar potential respects the custodial symmetry.
}
$m_{A^0,h_N^-}^2\sim \mu_N^2+\lambda_2v^2,~m_{h_N^0}^2\sim \mu_N^2+(\lambda_2-\lambda_3)v^2$.
Therefore, if we take the parameters by $\mu_N^2\ll v^2$ and $\lambda_2-\lambda_3={\cal O}(10^{-6})$ with $\lambda_{2,3}={\cal O}(1)$, desirable spectrum such as $m_{h_N^0}={\cal O}(1-10)~{\rm MeV}$ and $m_{A^0,h_N^-}\gtrsim 100~{\rm GeV}$ can be obtained without conflicting with the custodial symmetry.%
\footnote{In the triplet Higgs model, the mass splittings in the triplet Higgs multiplet leads 
to the custodial symmetry breaking. Thus, we cannot obtain a light particle with a mass 
in the MeV range while keeping other modes such as the charged Higgs bosons in the $O(100)$\,GeV range 
without conflicting with the custodial symmetry. 
The same problem arises in the Majoron model appearing from the triplet Higgs boson~\cite{Gelmini:1980re}.
}

It should be commented that the above assumption $\mu_N^2 \ll v^2$ is important for two reasons.
First, if $\mu_N^2 = O(v^2)$,  $ h_N^0$ in the MeV range is achieved by a cancellation between
two contributions, $\mu_N^2$ and $(\lambda_2 - \lambda_3)v^2$. 
In such a case, the vacuum at $h_N^0 = 0$ becomes unstable for a slightly larger field value of $h^0>v$
due to the negative value of $\lambda_3 - \lambda_2$. 
In order to avoid the instability, we need to assume $\mu_N^2 \ll v^2$ so that the lightness of $h_N^0$ is
achieved by the small but a positive value of ($\lambda_3 - \lambda_2$).
The second reason of this assumption is the suppression
of the invisible decay of the observed Higgs boson into a pair of $h_N^0$.
Under the assumption of $\mu_N^2 \ll v^2$, the value of ($\lambda_3-\lambda_2$) is inevitably small.
Thus, by remembering that  the branching ratio of the mode into a pair of $h_N^0$ is proportional 
to ($\lambda_3-\lambda_2$),
the lightness of the $h_N^0$ automatically guarantees the small branching ratio to 
a pair of $h_N^0$ under the assumption of $\mu_N^2 \ll v^2$.

Once we obtained the above mass splitting in the neutrinophilic Higgs doublet,
we obtain the effective theory of $h_N^0$ and the left-handed neutrinos,
\begin{eqnarray}
{\cal L}_{\rm eff} \simeq \frac{gyv}{M}h_N^0 \nu_L\nu_L\ ,
\end{eqnarray}
which realizes the model discussed in the previous section by identifying 
\begin{eqnarray}
g^{\rm eff} = \frac{gyv}{M}, \quad s = h_N^0\ .
\end{eqnarray}
By assuming $M= O(1)$\,TeV and $g = y = O(1)$, for example, we achieve 
the effective theory with $g^{\rm eff} = O(0.1)$.

The experimental limits on charged Higgs mass are given by using $t\to H^+ b$ for $m_{H^+}<m_t$ and $H^+\to \tau \nu$ for $m_{H^+}>m_t$ by $H^+$ production via third generation quarks at the LHC~\cite{ATLAS-CONF-2013-090}. 
However, $h_N$ does not couple to quarks in the model, and thus, $h_N^-$ is free from the limit.
So only the LEP constrains $h_N^-$ by $e^+e^-\to H^+H^-\to\tau\tau\nu\nu$, and the exclusion limit is $m_{H^+}\gtrsim 100~{\rm GeV}$ by imposing $Br(H^+\to\tau\nu)=1$~\cite{Abbiendi:2013hk}.
The CP-odd Higgs is still free from any experimental observation since it only couples with neutrino 
as long as it is heavier than the $Z$-boson.
Lepton flavor violation is also affected by the charged Higgs such as $\mu\to e\gamma$ induced by the effective operator $m_\mu(g^2/\Lambda^2)\bar\mu_R\sigma^{\mu\nu}e_LF_{\mu\nu}$ where $\Lambda$ is a cutoff scale.
Experimental limit is given by $Br(\mu\to e\gamma)\lesssim 10^{-13}$~\cite{Adam:2013mnn} which reads to $\Lambda\gtrsim{\cal O}(100)~{\rm GeV}$ if we take $g={\cal O}(0.1)$ and a loop factor is considered~\cite{deGouvea:2013zba}.

A crucial experimental limit is for the coupling among $h_N^0$ and neutrinos from the rare meson decay rates 
emitting $h_N^0$.
In particular, the null observations of $\pi/K \to l\nu_{l'}h_N^0$ put stringent constraints 
on the coupling $g$~\cite{Barger:1981vd,Lessa:2007up}.
Hereafter, we denote $g_{ab}^{\rm eff}h_N^0\bar\nu_a\nu_b~(a,b=e,\mu,\tau)$ as the flavor basis, 
and the coupling is converted into $g^{\rm eff}_{ij}=(U_{\rm PMNS}^\dagger)_{ia} g_{ab}^{\rm eff} (U_{\rm PMNS})_{bj}~(i,j=1,2,3)$ in the mass basis using the unitary PMNS matrix $U_{\rm PMNS}$:
\begin{eqnarray}
	U_{\rm PMNS} &=&
	{\footnotesize
	\left[
	\begin{array}{ccc}
		1 & 0 & 0 \\
		0 & c_{23} & s_{23} \\
		0 & -s_{23} & c_{23}
	\end{array}
	\right]
	\left[
	\begin{array}{ccc}
		c_{13} & 0 & s_{13}e^{-i\delta} \\
		0 & 1 & 0 \\
		-s_{13}e^{i\delta} & 0 & c_{13}
	\end{array}
	\right]
	\left[
	\begin{array}{ccc}
		c_{12} & s_{12} & 0 \\
		-s_{12} & c_{12} & 0 \\
		0 & 0 & 1 
	\end{array}
	\right]
	} \nonumber\\
	&& \times
	{\footnotesize
	\left[
	\begin{array}{ccc}
		e^{-\alpha_1/2} & 0 & 0 \\
		0 & e^{i\alpha_2/2} & 0 \\
		0 & 0 & 1 
	\end{array}
	\right]\ .
	}
\end{eqnarray}
Here, $s_{ij}\equiv\sin(\theta_{ij}),~c_{ij}\equiv\cos(\theta_{ij})$, and $\delta$ and $\alpha_i$ are Dirac and Majorana phases, respectively, and we take $\delta=\alpha_1=\alpha_2=0,~s_{12}^2=0.31,~s_{23}^2=0.51$ and $s_{13}^2=0.023$ in our analysis for simplicity.
In the flavor basis, the constraints from the rare meson decays put limit on $g_{ab}^{\rm eff}$~\cite{Lessa:2007up} 
\begin{equation}
	\sum_{l=e,\mu,\tau}|g^{\rm eff}_{el}|^2<5.5\times10^{-6},~\sum_{l=e,\mu,\tau}|g^{\rm eff}_{\mu l}|^2<4.5\times10^{-5}~{\rm and}~\sum_{l=e,\mu,\tau}|g^{\rm eff}_{\tau l}|^2<3.2\ .
	\label{eq:meson}
\end{equation}

As we have discussed in the previous section, we need to assume $g^{eff} = O(0.1$--$1)$ to 
obtain a short enough MFP for the neutrino flux from the sources inside our galaxy.
Due to the above constraints in Eq.\,(\ref{eq:meson}), the only allowed coupling of $O(0.1$-$1)$ 
is $g^{\rm eff}_{\tau\tau}$ in the flavor basis.
It should be noted though that in the mass basis, $g^{\rm eff}_{\tau\tau}$ leads to $O(1$--$0.1)$ 
couplings between three neutrinos in the mass basis according to the PNMS matrix.
If the sources of the neutrinos are the extra-galactic ones, on the other hand, the couplings
of $g^{eff} = O(0.01)$ are large enough to achieve the short MFP, which can easily evade
the constraints from the rare meson decay.

Now, let us calculate the resultant neutrino spectrum by assuming 
a single power-law flux at the neutrinos sources.
The number of neutrinos reaching to the Earth is approximately estimated by\footnote{To be more accurate, there exists other contributions such as the expansion effect of the universe and secondary neutrino scattering. However, in our case, the flight distance of neutrinos is small enough not to be affected by the expansion of the universe. For the secondary neutrino scattering effect, inelastic scattering is not sufficient since the SM cross section is negligible in the resonance region. In elastic scattering case, scattered neutrinos settle where their energy is around $E_\nu/2$, however, this contribution is also negligible and does not change our result where Eq.\,(\ref{eq:absorption}) is utilized.}
\begin{eqnarray}
	\frac{dN_\nu}{dL}(E_\nu,z)&\simeq&-\frac{N_\nu(E_\nu,z)}{\lambda(E_\nu)},
	\label{eq:absorption}
\end{eqnarray}
where $L$ is the length of the neutrino traveling path defined by 
\begin{eqnarray}
L=\frac{c}{H_0}\int dz(\Omega_m(1+z)^3+\Omega_\Lambda)^{-1/2}~{\rm Mpc}\ ,  
\end{eqnarray}
where $z$ denotes the redshift parameter, $c=3\times10^5~{\rm km/s},~H_0=100 h~{\rm km/s/Mpc}$, and 
$\Omega_m$ and $\Omega_\Lambda$ are energy densities of matter and dark energy, respectively.
In our analysis, we use $h=0.67,~\Omega_m=0.32$ and $\Omega_\Lambda=0.68$~\cite{Ade:2013zuv}.

\begin{center}
	\begin{figure}
		\includegraphics[scale=.55]{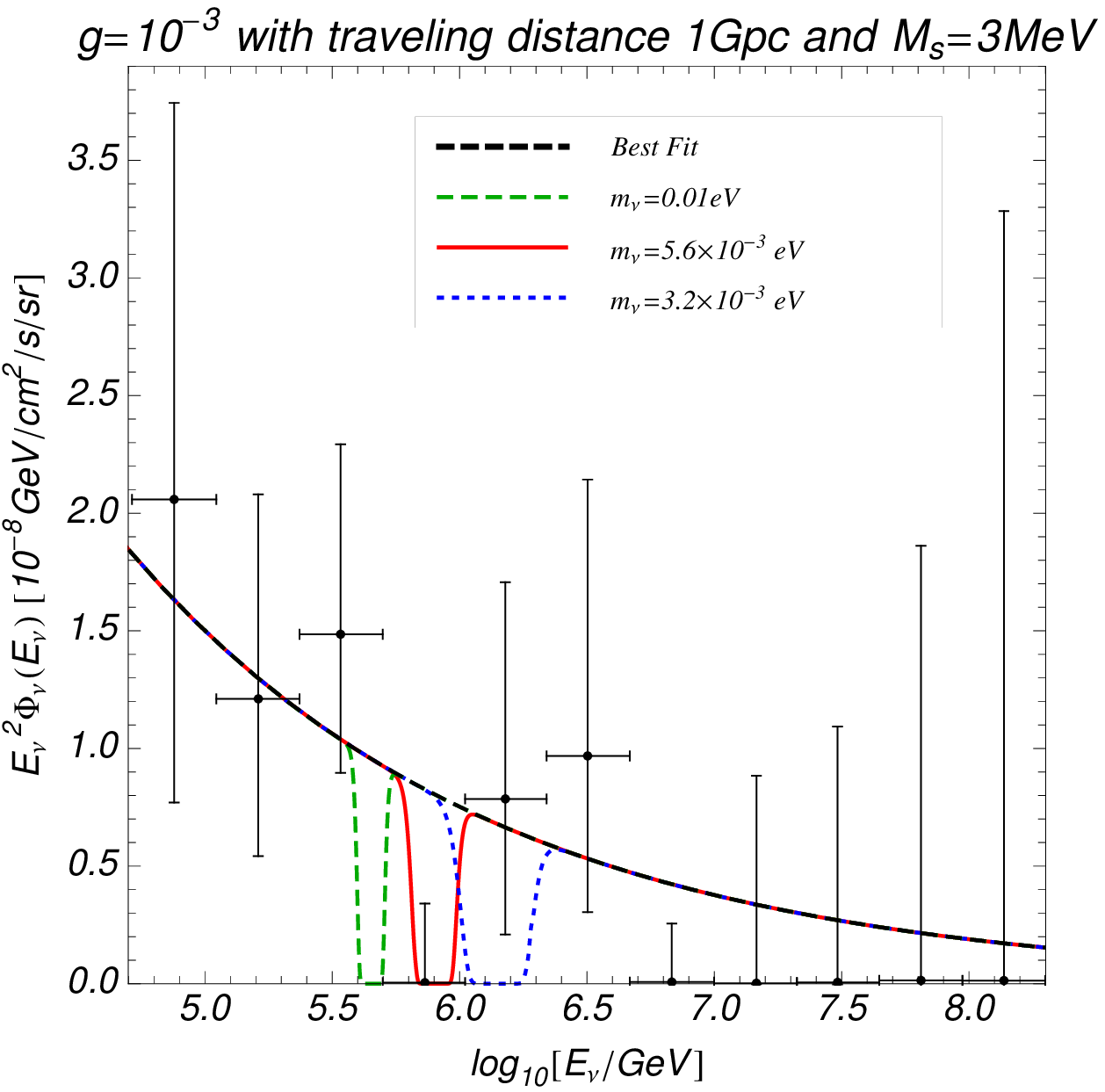}
		\includegraphics[scale=.55]{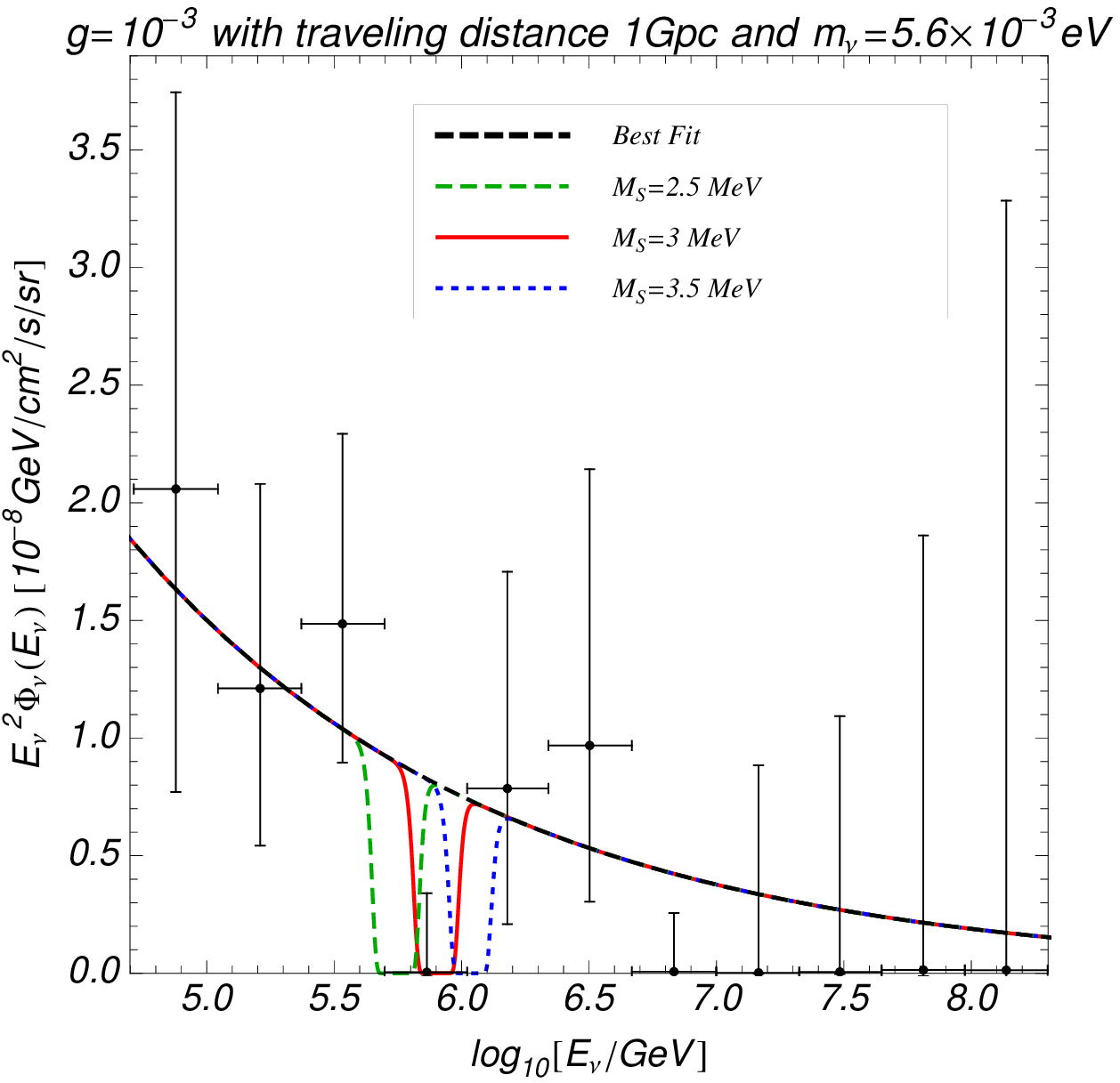}
		\caption{\sl\small Absorption line with the sample parameters assuming the source distance to be $O(1)$\,Gpc.
		{\bf Left panel:} The neutrino mass dependences. 
		The black dots with error bars are observed data, and the best-fit power law is 
		$E_\nu^2\Phi_\nu(E_\nu)=1.5\times10^{-8}(E_\nu/100{\rm TeV})^{-0.3}~{\rm GeV/cm^2/s/sr}$
		~\cite{Aartsen:2014gkd}. {\bf Right panel:} The dependence on the resonance mass.
		In both figures, we assumed $g \equiv g^{\rm eff}_{11} = O(10^{-3})$.
		}
		\label{fig3}
	\end{figure}
\end{center}

In Fig.~\ref{fig3}, we show some examples of the neutrino spectrum for the extra-galactic sources
locating at the distance of $1$\,Gpc.
The figure shows that the absorption line can be achieved for $g^{\rm eff}_{11} = 10^{-3}$,
which easily satisfies the constraints from the rare decay in Eq.\,(\ref{eq:meson}).
The left panel of the figure shows the neutrino mass dependence of the absorption line, where we take the neutrino mass as a free parameter, and focus on the dominant contribution to the absorption process.
From the figure, we find that the neutrino mass about $m_{\nu}\simeq 5.6\times 10^{-3}$\,eV
provides a nice fit to the null regions of the IceCube flux.
Interestingly, this mass is close to the square root of the squared mass differences of the
first two neutrinos in the normal hierarchy, $\Delta m_{21}^2\simeq 7.6\times10^{-5}~{\rm eV^2}$~\cite{Yao:2006px}.
Thus, the result favors the neutrino mass spectrum in which the first two neutrinos 
are rather degenerated.
The right panel shows the dependence on the resonance mass.
The figure shows that the nice fit is achieved for $M_s \simeq 3$\,MeV.

\begin{center}
	\begin{figure}
	\begin{center}
		\includegraphics[scale=.55]{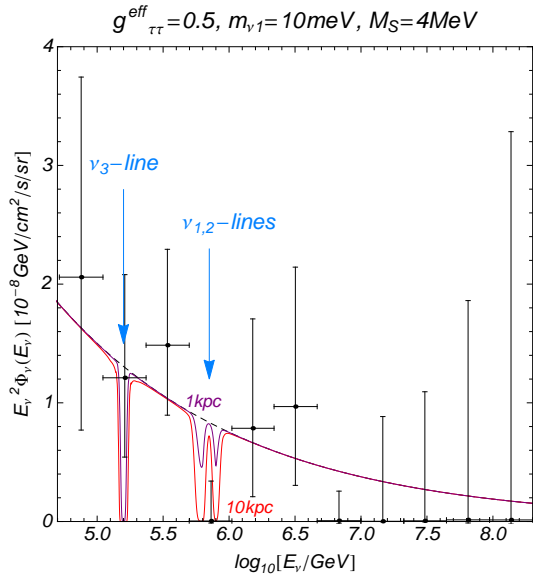}
		\caption{\sl\small Absorption line with the sample parameters assuming the source distance to be $O(10)$\,kpc.
		Here, we have taken $g_{\tau\tau}^{\rm eff} = 0.5$.
		}
		\label{fig4}
		\end{center}
	\end{figure}
\end{center}

When the neutrino sources are inside our galaxies with the distance of $O(10)$\,kpc, on the other hand,
we need to have $g^{eff}_{\tau\tau} = O(1)$. 
In this case, the meson decay constraints only allow $g^{\rm eff}_{\tau\tau} = O(1)$ in the flavor basis.
The coupling constant in the mass basis is, on the other hand, determined according to the PNMS
matrix with $g^{\rm eff}_{\tau\tau} = O(1)$.
As a result, the ratios between the coupling constants are fixed by the PNMS matrix
which leads to non-trivial relation between the absorption lines made by the three 
neutrinos.
In Fig.~\ref{fig4}, we show an example of the neutrino spectrum for $g_{\tau\tau}^{\rm eff} = 0.5$.
The figure shows that the spectrum has not only the broad absorption lines by the first two neutrinos
but also a sharp line by the third neutrino.
Here, again, the degenerated first two neutrinos are favored.
The detailed observation of the neutrino spectrum is required to test the existence of such 
multiple absorption lines in the neutrino spectrum.%
\footnote{Such multiple absorption lines are also possible for the extra-galactic neutrino sources
depending on the structure of the Yukawa couplings.}

\subsection{Another model}
Finally, let us discuss another possibility to induce the C$\nu$B absorption line at the sub-PeV scale.
As previously mentioned, the high energy neutrinos produced by astrophysical sources are mostly left-handed
even if the neutrinos are the Dirac type, and the chirality flip hardly takes place as they 
travel because its energy is much higher that the mass.
Therefore, it is simple to assume that the resonances appear in the collisions between the left-handed neutrinos.

When the neutrinos pass through the magnetic field, however, 
the chirality flip is potentially possible since the neutrinos have a finite magnetic momentum, $\mu_\nu$.
In the magnetic fields $B$, the Larmor frequency of the neutrino is given by $B \mu_\nu$, and hence,
the Dirac neutrinos flip their chirality when the travel time is longer than the Larmor frequency.%
\footnote{In the rest frame of the injecting neutrino, the travel time is suppressed by 
a large Lorentz boost factor, where the Larmor frequency is enhanced by the boost magnetic field
in the rest frame.
}
Thus, once the chirality flip occurs due to a strong magnetic field, 
the neutrino absorption can be achieved by the resonance appearing in the collisions between the right-handed neutrinos,
\begin{eqnarray}
{\cal L} = g s \bar{N_R} \bar{N_R}\ ,
\end{eqnarray}
in the case of the Dirac neutrino.  
The required masses of the resonance and the size of the coupling to obtain the visible absorption line are the similar
to the results in the previous section.
It should be noted that the size of the coupling $g$ is hardly constrained by any other experiments
including the rare meson decay.

Unfortunately, however, the neutrino magnetic moment predicted in the SM is very small,
\begin{eqnarray}
\mu_\nu \simeq 3\times 10^{-19} \left(\frac{m_\nu}{1\,{\rm eV}}\right) \mu_B\ , 
\end{eqnarray}
where $\mu_B\equiv e/(2m_e)\simeq 0.6\times 10^{-13}$GeV/T is the Bohr magneton.
Therefore the  necessary distance for the chirality flip is very long;
\begin{eqnarray}
L_{\rm cf} = \pi / \mu_\nu B \simeq  10\left(\frac{0.1\,{\rm eV}}{m_\nu}\right)\left(\frac{\mu G}{B}\right)\,{\rm Gpc}\ .
\end{eqnarray}
Thus, it is difficult to flip the chirality by the galactic magnetic field $B\simeq 10^{-6}$\,G~\cite{Barbieri:1988nh,Vogel:1989iv}.
As a result, in order for the chirality flip takes place, we need a new physics which enhances the 
neutrino magnetic moment significantly (see e.g. Ref.~\cite{Aboubrahim:2013yfa}).
For example, if we assume the current experimental upper limit on the neutrino magnetic field,
 $\mu_\nu<5.4\times10^{-11}\mu_B$~\cite{Arpesella:2008mt}, 
 the chirality flip is possible within the traveling dietance of
 ${\cal O}(1)~{\rm kpc}$ under the the galactic magnetic field $B\simeq 10^{-6}~{\rm G}$.%
\footnote{
If the neutrinos are accelerated at the PWN surrounding a very strong magnetic field $B\sim 10^{12}$\,G,
the required enhancement of the neutrino magnetic moment can be smaller, although
we do not pursue this possibility any more in this paper.}

\section{Summary and Discussion}
\label{sec:summary}


In this paper, we have discussed the possibility whether the 
null-event region around the sub-PeV scale in the neutrino spectrum observed at the IceCube
experiment can be interpreted as an absorption line by the C$\nu$B 
in the power-law spectrum. 
To achieve such a possibility, we proposed two viable models where the MeV resonance
appears in the neutrino-neutrino interactions. 
For the models with Majorana neutrinos, we found that the resonance is embedded in the
neutrinophilic doublet boson which will be tested by future collider experiments.
For the models with the Dirac neutrinos, we found that the resonance appearing 
in the interaction of the right-handed neutrinos is also a possibility, although 
we need an enhancement of the neutrino magnetic moment to flip the chirality
of the neutrinos 
during the flight to hit the resonance.
Such an enhanced neutrino magnetic momentum requires an additional new physics
beyond the SM, which will also be tested by future collider experiments.

It should be noted that the shape of the absorption line depends not only the mass
of the new resonance but also on the neutrino masses.
Thus, in principle, it is possible to extract the masses of the neutrinos by investigating
the absorption lines in the neutrino spectrum, although it requires very high energy resolution.
The identification of the astrophysical sources of the high energy neutrinos is also crucial
to determine the absorption line, since it depends on the relative magnitude between 
the MFP and the distance to the neutrino source from the Earth.

Finally let us comment on an implication for cosmology.
Non-standard neutrino interactions can affect the CMB power spectrum and/or the structure formation 
of the universe since it might change the decoupling temperature of the neutrinos  
and/or the neutrino free-streaming scale.
Interestingly, the recent CMB analysis~\cite{Cyr-Racine:2013jua} 
reported a slight preference for an additional neutrino interactions with the magnitude of 
$g^2/M_s^2\simeq 1/(10~{\rm MeV})^2$ which is surprisingly close to the ones we are assuming.
Since the conclusion has a prior dependence~\cite{Archidiacono:2013dua}, it is premature to say that the existence of the
non-standard neutrino interactions are supported by the CMB observation.
However, such cosmological observations are expected to provide significant synergy of the IceCube experiment 
in future studies.

\section{Acknowledgments}
We thank T.~Fujita for useful discussions.
We also thank S.~Yoshida for useful discussion on the neutrino 
absorption in the SM.
This work is supported by Grant-in-Aid for Scientific research 
from the Ministry of Education, Science, Sports, and Culture (MEXT), Japan, No. 24740151 
and 25105011 (M.I.), from the Japan Society for the Promotion of Science (JSPS), No. 26287039 (M.I.).
%
This work is also supported by World Premier International Research Center Initiative (WPI Initiative), MEXT, Japan.

\bibliographystyle{h-physrev}
\bibliography{Reference}

\begin{thebibliography}{10}

\bibitem{Achterberg:2006md}
IceCube Collaboration, A.~Achterberg {\em et~al.},
\newblock Astropart.Phys. {\bf 26}, 155 (2006), astro-ph/0604450.

\bibitem{Aartsen:2014gkd}
IceCube Collaboration, M.~Aartsen {\em et~al.},
\newblock (2014), 1405.5303.

\bibitem{Lunardini:2000fy}
C.~Lunardini and A.~Y. Smirnov,
\newblock Phys.Rev. {\bf D64}, 073006 (2001), hep-ph/0012056.

\bibitem{Athar:2000yw}
H.~Athar, M.~Jezabek, and O.~Yasuda,
\newblock Phys.Rev. {\bf D62}, 103007 (2000), hep-ph/0005104.

\bibitem{Yoshida:1993pt}
S.~Yoshida and M.~Teshima,
\newblock Prog.Theor.Phys. {\bf 89}, 833 (1993).

\bibitem{Takami:2007pp}
H.~Takami, K.~Murase, S.~Nagataki, and K.~Sato,
\newblock Astropart.Phys. {\bf 31}, 201 (2009), 0704.0979,
\newblock and references therein.

\bibitem{Villante:2008qg}
F.~Villante and F.~Vissani,
\newblock Phys.Rev. {\bf D78}, 103007 (2008), 0807.4151,
\newblock and references therein.

\bibitem{Murase:2013ffa}
K.~Murase and K.~Ioka,
\newblock Phys.Rev.Lett. {\bf 111}, 121102 (2013), 1306.2274.

\bibitem{Bednarek:2003cv}
W.~Bednarek,
\newblock Astron.Astrophys. {\bf 407}, 1 (2003), astro-ph/0305430,
\newblock and references therein.

\bibitem{Asano:2014nba}
K.~Asano and P.~Mszaros,
\newblock Astrophys.J. {\bf 785}, 54 (2014), 1402.6057,
\newblock and references therein.

\bibitem{Dado:2014mea}
S.~Dado and A.~Dar,
\newblock (2014), 1405.5487,
\newblock and references therein.

\bibitem{Stecker:2013fxa}
F.~W. Stecker,
\newblock Phys.Rev. {\bf D88}, 047301 (2013), 1305.7404,
\newblock and references therein.

\bibitem{Tamborra:2014xia}
I.~Tamborra, S.~Ando, and K.~Murase,
\newblock (2014), 1404.1189,
\newblock and references therein.

\bibitem{Learned:2000sw}
J.~Learned and K.~Mannheim,
\newblock Ann.Rev.Nucl.Part.Sci. {\bf 50}, 679 (2000).

\bibitem{Dermer:2006xt}
C.~D. Dermer,
\newblock J.Phys.Conf.Ser. {\bf 60}, 8 (2007), astro-ph/0611191.

\bibitem{Cholis:2012kq}
I.~Cholis and D.~Hooper,
\newblock JCAP {\bf 1306}, 030 (2013), 1211.1974.

\bibitem{Murase:2013rfa}
K.~Murase, M.~Ahlers, and B.~C. Lacki,
\newblock Phys.Rev. {\bf D88}, 121301 (2013), 1306.3417.

\bibitem{Anchordoqui:2013dnh}
L.~A. Anchordoqui {\em et~al.},
\newblock Journal of High Energy Astrophysics {\bf 1-2}, 1 (2014), 1312.6587.

\bibitem{Feldstein:2013kka}
B.~Feldstein, A.~Kusenko, S.~Matsumoto, and T.~T. Yanagida,
\newblock Phys.Rev. {\bf D88}, 015004 (2013), 1303.7320.

\bibitem{Ema:2013nda}
Y.~Ema, R.~Jinno, and T.~Moroi,
\newblock Phys.Lett. {\bf B733}, 120 (2014), 1312.3501.

\bibitem{Esmaili:2013gha}
A.~Esmaili and P.~D. Serpico,
\newblock JCAP {\bf 1311}, 054 (2013), 1308.1105.

\bibitem{Ng:2014pca}
K.~C.~Y. Ng and J.~F. Beacom,
\newblock (2014), 1404.2288.

\bibitem{Ioka:2014kca}
K.~Ioka and K.~Murase,
\newblock PTEP {\bf 2014}, 061E01 (2014), 1404.2279.

\bibitem{Zavala:2014dla}
J.~Zavala,
\newblock (2014), 1404.2932.

\bibitem{Chen:2013dza}
C.-Y. Chen, P.~S.~B. Dev, and A.~Soni,
\newblock Phys.Rev. {\bf D89}, 033012 (2014), 1309.1764.

\bibitem{Roulet:1992pz}
E.~Roulet,
\newblock Phys.Rev. {\bf D47}, 5247 (1993).

\bibitem{Weiler:1982qy}
T.~J. Weiler,
\newblock Phys.Rev.Lett. {\bf 49}, 234 (1982).

\bibitem{Weiler:1983xx}
T.~J. Weiler,
\newblock Astrophys.J. {\bf 285}, 495 (1984).

\bibitem{Weiler:1997sh}
T.~J. Weiler,
\newblock Astropart.Phys. {\bf 11}, 303 (1999), hep-ph/9710431.

\bibitem{Yoshida:1994ci}
S.~Yoshida,
\newblock Astropart.Phys. {\bf 2}, 187 (1994).

\bibitem{Yoshida:1996ie}
S.~Yoshida, H.-y. Dai, C.~C. Jui, and P.~Sommers,
\newblock Astrophys.J. {\bf 479}, 547 (1997), astro-ph/9608186.

\bibitem{Fodor:2001qy}
Z.~Fodor, S.~Katz, and A.~Ringwald,
\newblock Phys.Rev.Lett. {\bf 88}, 171101 (2002), hep-ph/0105064.

\bibitem{Eberle:2004ua}
B.~Eberle, A.~Ringwald, L.~Song, and T.~J. Weiler,
\newblock Phys.Rev. {\bf D70}, 023007 (2004), hep-ph/0401203.

\bibitem{Barenboim:2004di}
G.~Barenboim, O.~Mena~Requejo, and C.~Quigg,
\newblock Phys.Rev. {\bf D71}, 083002 (2005), hep-ph/0412122.

\bibitem{Kirilova:2014ipa}
D.~Kirilova,
\newblock (2014), 1407.1784,
\newblock and references therein.

\bibitem{Chikashige:1980ui}
Y.~Chikashige, R.~N. Mohapatra, and R.~Peccei,
\newblock Phys.Lett. {\bf B98}, 265 (1981).

\bibitem{Gelmini:1980re}
G.~Gelmini and M.~Roncadelli,
\newblock Phys.Lett. {\bf B99}, 411 (1981).

\bibitem{Ma:2000cc}
E.~Ma,
\newblock Phys.Rev.Lett. {\bf 86}, 2502 (2001), hep-ph/0011121.

\bibitem{Haba:2011ra}
N.~Haba and O.~Seto,
\newblock Prog.Theor.Phys. {\bf 125}, 1155 (2011), 1102.2889.

\bibitem{Chanowitz:1985ug}
M.~S. Chanowitz and M.~Golden,
\newblock Phys.Lett. {\bf B165}, 105 (1985).

\bibitem{Deppisch:2004fa}
F.~Deppisch and J.~Valle,
\newblock Phys.Rev. {\bf D72}, 036001 (2005), hep-ph/0406040.

\bibitem{ATLAS-CONF-2013-090}
CERN Report No. ATLAS-CONF-2013-090, 2013 (unpublished).

\bibitem{Abbiendi:2013hk}
ALEPH, DELPHI, L3, OPAL, LEP, G.~Abbiendi {\em et~al.},
\newblock Eur.Phys.J. {\bf C73}, 2463 (2013), 1301.6065.

\bibitem{Adam:2013mnn}
MEG Collaboration, J.~Adam {\em et~al.},
\newblock Phys.Rev.Lett. {\bf 110}, 201801 (2013), 1303.0754.

\bibitem{deGouvea:2013zba}
A.~de~Gouvea and P.~Vogel,
\newblock Prog.Part.Nucl.Phys. {\bf 71}, 75 (2013), 1303.4097.

\bibitem{Barger:1981vd}
V.~D. Barger, W.-Y. Keung, and S.~Pakvasa,
\newblock Phys.Rev. {\bf D25}, 907 (1982).

\bibitem{Lessa:2007up}
A.~Lessa and O.~Peres,
\newblock Phys.Rev. {\bf D75}, 094001 (2007), hep-ph/0701068.

\bibitem{Ade:2013zuv}
Planck Collaboration, P.~Ade {\em et~al.},
\newblock (2013), 1303.5076.

\bibitem{Yao:2006px}
Particle Data Group, W.~Yao {\em et~al.},
\newblock J.Phys. {\bf G33}, 1 (2006).

\bibitem{Barbieri:1988nh}
R.~Barbieri and R.~N. Mohapatra,
\newblock Phys.Rev.Lett. {\bf 61}, 27 (1988).

\bibitem{Vogel:1989iv}
P.~Vogel and J.~Engel,
\newblock Phys.Rev. {\bf D39}, 3378 (1989).

\bibitem{Aboubrahim:2013yfa}
A.~Aboubrahim, T.~Ibrahim, A.~Itani, and P.~Nath,
\newblock Phys.Rev. {\bf D89}, 055009 (2014), 1312.2505.

\bibitem{Arpesella:2008mt}
Borexino Collaboration, C.~Arpesella {\em et~al.},
\newblock Phys.Rev.Lett. {\bf 101}, 091302 (2008), 0805.3843.

\bibitem{Cyr-Racine:2013jua}
F.-Y. Cyr-Racine and K.~Sigurdson,
\newblock (2013), 1306.1536.

\bibitem{Archidiacono:2013dua}
M.~Archidiacono and S.~Hannestad,
\newblock (2013), 1311.3873.

\end{thebibliography}

\end{document}